\documentclass[12pt]{iopart}

\usepackage{graphicx,amssymb}

\newcommand{\be}{\begin{equation}}
\newcommand{\ee}{\end{equation}}
\newcommand{\bea}{\begin{eqnarray}}
\newcommand{\eea}{\end{eqnarray}}
\renewcommand\({\left(}
\renewcommand\){\right)}

\newcommand{\RT}{{\scriptscriptstyle \rm RT}}
\newcommand{\m}{m}

\providecommand{\eref}[1]{(\ref{#1})}
\providecommand{\Or}{{\mathcal O}}
\providecommand{\fl}{}
\providecommand{\rmi}{{\rm i}}
\providecommand{\ack}{\subsubsection*{Acknowledgment}}
\providecommand{\fl}{}

\begin{document}
\hfill DESY 08-049

\title{Racetrack Inflation and Cosmic Strings}

\author{Ph~Brax$^1$, C~van~de~Bruck$^2$, A~C~Davis$^3$, S~C~Davis$^3$,
R~Jeannerot$^4$ and M~Postma$^{5,6}$}

\address{ ${}^1$ 
Institut de Physique Th\'eorique,
CEA, IPhT, F-91191 Gif-sur-Yvette, France.
CNRS, URA 2306, F-91191 Gif-sur-Yvette, France
}
\address{ ${}^2$ Department of Applied Mathematics, School of
  Mathematics and Statistics, 
University of Sheffield, Houndsfield Road, Sheffield, S3 7RH, UK}
\address{${}^3$ DAMTP, Centre for Mathematical Sciences,
University of Cambridge, Wilberforce Road, Cambridge, CB3 0WA, UK}
\address{${}^4$ Instituut-Lorentz for Theoretical Physics, Postbus 9506,
 2300 RA Leiden, The Netherlands}
\address{${}^5$ DESY, Notkestra\ss e 85, 22607 Hamburg, Germany}
\address{${}^6$ Nikhef, Kruislaan 409, 1098 SJ Amsterdam, The
Netherlands}

\eads{\mailto{brax@spht.saclay.cea.fr}, \mailto{c.vandebruck@shef.ac.uk},
\mailto{A.C.Davis@damtp.cam.ac.uk}, \mailto{S.C.Davis@damtp.cam.ac.uk},
\mailto{jeannero@lorentz.leidenuniv.nl}, \mailto{postma@mail.desy.de}}

\begin{abstract}
We consider the coupling of racetrack inflation to matter fields
as realised in the D3/D7 brane system. In particular, we
investigate the possibility of cosmic string formation in this
system. We find that string formation before or at the onset of racetrack
inflation is possible, but they  are then inflated away.
Furthermore, string formation at the end of inflation is prevented
by the presence of the moduli sector. As a consequence, no strings
survive racetrack inflation.
\end{abstract}


\section{Introduction}
Racetrack inflation is a promising inflationary model, which can
be realised within string theory \cite{rt1,rt2}. In this scenario,
inflation is driven by K\"ahler moduli fields. The original string
theory scenario is based on the KKLT mechanism for moduli fixing,
extending it to include a racetrack-type superpotential, i.e.\ the
superpotential contains more than one exponential of the K\"ahler
modulus field \cite{rt1}. In a later development, based on an
explicit compactification of type IIB string theory, racetrack
inflation with two complex K\"ahler moduli has been considered
\cite{rt2}. In these papers, the KKLT set-up has been used, in
which a flux potential stabilises the dilaton and the complex
structure moduli. A non-trivial potential for the K\"ahler moduli is
generated by non-perturbative effects (for example via gaugino
condensation). This alone results in an Anti-de Sitter
(AdS) vacuum, which is then uplifted by the presence of anti-branes,
which break supersymmetry explicitly. In \cite{raceD} a racetrack
model was constructed, in which the uplifting is obtained by
$D$-terms, as suggested in \cite{Dterm1,Dterm2}.

In all cases the possible presence of additional fields
during racetrack inflation has been ignored (apart from
\cite{raceD}, where additional meson fields on D7 branes have been
included). In supergravity, interactions between the K\"ahler  modulus field
and other fields are inevitable and one may wonder whether the
dynamics of the modulus field are  slightly modified. With the meson
field considered in \cite{raceD}, the features of racetrack
inflation were not significantly altered. Additionally,
topological defects, such as cosmic strings, may form during or
after inflation. In this paper, we extend the racetrack
inflation scenario and study the impact of such additional fields
on inflation. The model we consider is inspired by the D3/D7 system in
type IIB string theory compactified on $K3\times T^2/Z_2$~\cite{D3D7}. 
On top of a K\"ahler modulus field $T$, the system contains
a neutral field $\phi$, describing the interbrane distance and two
charged fields $\phi^\pm$, describing open strings stretching
between the D3 and the D7 brane. This construction is the stringy
analogue of $D$-term hybrid inflation. The main difference we
introduce here is that inflation is due to the racetrack sector
and is not driven by the interbrane potential. Cosmic strings are
formed when the charged fields condense. We investigate whether the
formation of these strings is still possible in this scenario.

The condensation scale depends on a Fayet-Iliopoulos (FI) term which
originates from a non-trivial flux on D7 branes. In the low energy
supergravity description, such FI terms are either constant when a
$U(1)_R$ gauged R-symmetry is present or field dependent when the
$U(1)$ gauge symmetry is pseudo-anomalous. Here the structure of the
superpotential, after the dilaton and the complex structure moduli have
been stabilised by fluxes, prevents the existence of the R-symmetry.
Pseudo-anomalous symmetries with a field dependent FI term exist and
can be used to uplift the potential~\cite{raceD}. $D$-term cosmic
strings necessitate the introduction of  either two FI terms, one for
uplifting and one for condensation\footnote{With only one FI term, the
uplifting term is removed by the charged field condensation.} or a
single FI term together with an uplifting anti-brane. In fact, we will
see that even when no FI term is introduced the condensation of the
charged fields can be triggered by the $F$-term potential. This is a
purely supergravity effect; the U(1) symmetry breaking is induced by
the moduli sector.  If this occurs the $F$-term strings that form during
the phase transition have a tension that depends on the vacuum
expectation values of the moduli fields.  In all cases, we find that
either strings are inflated away when created before or during
inflation, or they are dynamically prevented from forming in the first
place.

The paper is organised as follows. In Section 2 we describe a simple
toy model in which cosmic strings form, and whose string tension is field
dependent. In Section 3 we review the setup of racetrack inflation. In
Section 4 we couple the racetrack model to matter fields and study the
dynamics of the fields during inflation. We discuss the conditions
under which cosmic strings form. We discuss and summarise our findings
in Section 5.


\section{Field dependent $F$-term strings}

In this section, we describe a simple situation where field
dependent cosmic strings may form.  By ``field dependent'' we mean
that the string tension depends on the vacuum expectation values
(VEVs) of other fields in the theory.  Cosmic strings form during
a phase transition if the vacuum manifold has non-contractible
loops, as is the case in a U(1) symmetry breaking.

A U(1)-breaking phase transition will occur in a supersymmetric theory with
the superpotential and $D$-term potentials
\begin{equation}
W= \lambda \Phi (\Phi^+\Phi^- -x^2) \, ,
\qquad
V_D = \frac{g^2}{2} \(|\Phi^+|^2 - |\Phi^-|^2 - \xi\)^2 \, ,
\end{equation}
where $\lambda$ is a coupling constant, $g$ the U(1) gauge coupling, $x$ a
mass parameter, and $\xi$ a Fayet-Iliopolous term. The fields
$\Phi^\pm$ are oppositely charged, while the third field $\Phi$ is neutral.
If the value of $\Phi$ is below some critical value, the charged
fields will have a tachyonic instability and will condense.  For a
$F$-term driven model with no FI term, $\Phi^+\Phi^-=x^2$ in the true vacuum.
For $D$-term driven models $x = 0$, and the global minimum has $\Phi^- =0$ and
$\Phi^+ = \sqrt{\xi}$.  In both cases the U(1) symmetry breaking
produces cosmic strings.  We will refer to these
two types of strings as $F$-term and $D$-term strings\footnote{Not to
be confused with the stringy usage of F- and D-strings, which refer to
fundamental and Dirichlet strings respectively.}.

We note that apart from symmetry breaking, the above theory can also
give rise to hybrid inflation. In this paper we will be
mainly be interested in models in which inflation is produced by a different
sector of the full theory. However some features of supersymmetric
hybrid inflation will be relevant to our analysis, so we will briefly
review it here. In hybrid inflation models $\Phi$ is the inflaton.
For $\langle \Phi \rangle$ larger than the critical value, the
potential has a valley of local minima where inflation takes
place. The U(1) gauge symmetry is unbroken and the ``waterfall''
fields $\Phi^\pm$ are zero. At tree level the inflationary direction is flat,
but it is lifted by loop corrections which induce the slow rolling of the
inflaton.  When $\Phi$ falls below the critical value, the phase
transition described above ends inflation, and cosmic strings form.

Finding a non-vanishing constant $x$-term in string theory has proved
to be difficult. Instead an FI term, and thus $D$-term symmetry breaking,
is easily realised by a D3/D7 system. This, and the corresponding
hybrid inflation model, is described in \cite{D3D7}.  When
moduli fields are introduced, the presence of a constant FI term is
forbidden\footnote{In supergravity a constant FI term can only be
  introduced if the theory is invariant under a gauged R-symmetry.
  However the constant $W_0$ in the KKLT moduli superpotential breaks
  this R-symmetry.}. A moduli dependent FI term is required instead,
leading to the formation of field dependent $D$-term strings. However,
as was shown in \cite{DtermHI}, in this set-up it is hard to combine
inflation and moduli stabilisation in a working model. In fact, the
problems of combining inflation with moduli stabilisation occur in a
wide range of models~\cite{infmod}, and are not
restricted to $D$-term hybrid inflation.

In the following we describe a set-up in which an effective $x$-term
is induced by the presence of the moduli sector.  This overcomes the
difficulty of finding a constant $x$-parameter. Just as in hybrid inflation
models we assume a shift-symmetric K\"ahler potential for the gauge
singlet \cite{shift}. In our set-up inflation is not driven by the
gauge singlet but by the moduli fields, implemented in the form of
racetrack inflation. In the remainder of this section we outline the
basic idea.  

Our set-up is loosely based  on the  D3/D7-matter system in type IIB
string theory compactified on $K3 \times T^2/Z_2$~\cite{D3D7}. 
Although we will not derive an exact
correspondence, we will say a bit more on this later on.  For now,
just consider a supergravity theory with the following super- and
K\"ahler potentials
\begin{equation}
W = \lambda \Phi \Phi^+\Phi^- \, , \qquad
K =  -\frac12(\Phi -\bar \Phi)^2 + |\Phi_+|^2 + |\Phi_-|^2 \, .
\end{equation}
Note that there is no constant $x$-term in the superpotential.
We embed this
model in a supersymmetry breaking ($m_{3/2}\ne 0$) and moduli
stabilised background.  The moduli scalar potential comprises $F$- and
uplifting terms such that $V_{\rm mod} \sim H_*^2$ during inflation and
$V_{\rm mod} \sim 0$ in the Minkowski vacuum after inflation (the
above relations are up to corrections coming from the matter sector).
Here we have defined $H_*$ as the Hubble rate during inflation. We
assume that the matter sector parameterised by $\Phi$ and $\Phi^\pm$
is a small perturbation to the moduli dynamics, i.e.\ 
$\Phi,\ \Phi^\pm \ll 1$. In this limit, for $\Phi =0$, the potential reads
\begin{eqnarray}
\fl
V = & 
\frac{g^2}{2} \(\vert\Phi^+\vert^2 -\vert\Phi^-\vert^2\)^2
+ V^F_{\rm mod} + V_{\rm up} 
+ \e^{K_{\rm mod}} \lambda^2 |\Phi^+|^2 |\Phi^-|^2
\nonumber  \\ \fl & {}
+\(\frac{V_{\rm mod}^F}{2} +  m_{3/2}^2\)  \(|\Phi^+|^2 +|\Phi^-|^2\)^2
+ (V_{\rm mod}^F + m_{3/2}^2)  \(|\Phi^+|^2 +|\Phi^-|^2\)
\end{eqnarray}
with $K_{\rm mod}$ the K\"ahler potential
of the moduli sector, and $m_{3/2}$ is the gravitino mass (defined in
the absence of the matter fields). The moduli stabilisation potential is 
$V_{\rm  mod} = V^F_{\rm mod} + V_{\rm up}$, where $V^F_{\rm mod}$ and
$V_{\rm up}$ are respectively the $F$- and uplifting terms. Note that
$V^F_{\rm mod} <0$. For $2m_{3/2}^2 > -V^F_{\rm mod} > m_{3/2}^2$, the
final three terms of the above potential resemble the usual symmetry
breaking terms that are generated by a non-zero $x$ or FI term.  Since
these terms arise in the $F$-term potential, it appears that an
effective field-dependent $x$-term is generated.

The dynamics of the model are reminiscent of what happens in hybrid
inflation (remember that in our set-up, the moduli sector is
responsible for inflation, and not the matter fields).  The
$\phi = {\rm Re}(\Phi)$ direction is flat at tree  level, and only lifted by
loop corrections.  If the loop corrections are small, the $\phi$-field
is frozen during inflation.  The masses of the waterfall fields are
$\phi$-dependent.  If initially $\phi > \phi_c$ with $\phi_c$ some
critical value, the charged fields are minimised at $\Phi^\pm =0$ and
the U(1) symmetry is unbroken.  Some time after inflation, when $H
\sim \partial_\phi^2 V_{\rm loop}$, the $\phi$ field starts rolling
down its potential dropping below the critical value. In the usual hybrid
inflation this triggers the U(1) breaking phase transition.  In our
set-up the situation is more complicated, and as we will see whether
the waterfall fields actually condense at low energies depends on the
specifics of the moduli sector.

The complete model will be studied in section \ref{s:matter}, where both
the inflaton/moduli and the matter sectors are treated
dynamically.  As discussed above, this full treatment is needed to
determine whether cosmic strings can form.

\section{Racetrack Inflation}
\label{s:racetrack}

In this section we will briefly review racetrack inflation in a
supergravity setting.  The original racetrack model \cite{rt1} is
formulated in a flux compactification of type IIB string on a
Calabi-Yau space. In the low energy effective action, there is only
the volume modulus field $T$, with a no-scale K\"ahler potential
\be
K^\RT = -3 \log(T+\bar T) \, .
\label{KRT}
\ee
The superpotential is of the modified racetrack form
\be
W^\RT = W_0 + A \e^{-aT}  + B \e^{-b T} \, .
\label{WRT}
\ee
%
The constant $W_0$ arises from integrating out the stabilised
dilaton and the complex structure moduli.  The non-perturbative terms
come from gaugino condensation on D7 branes or from instanton effects; in
both cases the parameters $a,b$ depend on the specifics of the gauge
group.  In addition there is an uplifting term
\be
V_{\rm up} = \frac{E}{(T+\bar T)^n}
\label{up}
\ee
originating from an anti-D3 in the bulk ($n=3$) or in the throat
($n=2$).  The constant $E$ is tuned to get a Minkowski vacuum in the
minimum after inflation.

The full potential is a series of cosines in $Y$, where we defined 
$T = X + \rmi Y$.  Inflation takes place near a saddle point, which is
unstable in the $Y$-direction but stable in the $X$-direction (in
order that $X$ does not run off to infinity during inflation). The
overall scale of the potential is set by the WMAP normalisation and we
find that the Hubble parameter is $H_* \sim m_{3/2} \sim 10^{-8}$ with
$m_{3/2} = \e^{K/2} |W|$ the vacuum gravitino mass. This fixes the
constant term of the superpotential $W_0 \sim 10^{-4} - 10^{-5}$. The
spectral index $n_s \leq 0.95$, in good agreement with the latest WMAP
data \cite{WMAP}.

In the above set-up the uplifting term breaks SUSY explicitly.
This can be cured by using instead an uplifting $D$-term
\cite{raceD,Dterm1,Dterm2}; in this case additional meson fields
have to be introduced to make the potential gauge invariant.  In
the improved racetrack model \cite{rt2}, a set-up is discussed
with two K\"ahler moduli.  This all suggests that racetrack is
very robust, and does not depend on the details of the uplifting
or the superpotential (though at least two exponents are needed).
This is in line with the observation made in \cite{accidental}
that saddle points are ubiquitous in the string landscape.


\section{Cosmic Strings and Racetrack Inflation}
\label{s:matter}

In this section we discuss the dynamics of racetrack inflation in the presence of a matter sector. 
Our set-up are  inspired by the D3/D7 system described in
\cite{D3D7}, in the context of type IIB theory compactified on
$K3 \times T^2/Z_2$. Consider a D7 and D3 brane which are located
far from the gaugino condensation brane. Then the light D3/D7 matter
fields are uncharged under the moduli sector symmetries. In the
same vein we neglect the backreaction of the D3 on the geometry,
which makes the $A$ and $B$ in the superpotential on the gaugino D7 brane
dependent on the D3 position~\cite{bergbaumanncline}. We thus assume
that the racetrack fields couple only gravitationally to the
matter fields.

The fields in the matter D3/D7 sector are a neutral
field $\Phi$ representing the interbrane distance, and two
oppositely charged fields $\Phi^\pm$ corresponding to strings
stretching between the D3 and D7 branes.
Approximate translational invariance of the brane system, which is a
consequence of the background isometries, translates into a
shift symmetry for the $\Phi$ field in the K\"ahler.    For simplicity we
assume canonical kinetic terms for the charged fields\footnote{In a
  set-up with the D3 at a fixed point, or in the limit that the stack
  of D3s is heavy, we can treat the D7 as a probe brane.  Then the
  matter fields have unit modular weight, and in the small field
  limit can be expanded to get minimal kinetic terms at lowest
  order.}; we expect that more complicated K\"ahler potentials give
qualitatively similar results.

For the moduli sector we use the no-scale modulus with a modified
racetrack potential discussed in the previous section.  Note that
in the explicit example based on  $K3 \times T^2/Z_2$, there are
many more K\"ahler moduli \cite{trivedi}.  In particular, in addition to
the $K3$ volume modulus (that we are calling $T$), there is the
volume modulus of the torus.  We assume that all these additional
moduli are stabilised by instanton effects \cite{aspinwall}. If
the torus is stabilised with the same radius as the $K3$ manifold the
effective K\"ahler potential for $T$ is of the no-scale form
\cite{shift}. In any case, as we remarked in the previous section,
racetrack inflation is very robust.  In that spirit we can use the
racetrack model (\ref{KRT}, \ref{WRT},~\ref{up}) as a useful toy
model for a possibly more complex set-up.

The model we study  is then
\bea
K &=& K^\RT + K^{\rm m} \,= -3\log(T+\bar T) 
-\frac12(\Phi -\bar \Phi)^2 + |\Phi^+|^2 + |\Phi^-|^2 \, , \\
W &=& W^\RT + W^{\rm m}
= W_0 + A\e^{-aT} + B \e^{-bT} + \lambda \Phi \Phi^+ \Phi^- \, .
\label{W}
\eea
The scalar potential reads
\be \fl V =
\frac{\e^{K^{\rm m}}}{(T+\bar T)^3} \Bigg( \frac{(T+\bar T)^2}{3}|F_T|^2
+ \sum_i|F_{\Phi_i}|^2 - 3 |W|^2 \Bigg)
+ V_{\rm up} 
+ \frac{g^2}{2} \bigg(\vert \Phi_+\vert^2 -\vert\Phi_-\vert^2 \bigg)^2
\ee
where
\begin{equation}
F_T=F_T^\RT -\frac{3}{T+\bar T} \lambda \Phi\Phi_+\Phi_- \, ,\
\end{equation}
and
\bea
F_{\Phi^\pm} &=& \lambda\Phi^\mp \Phi +\bar \Phi^\pm
(W^\RT +\lambda \Phi\Phi^+\Phi^-) \, , \\
F_\Phi&=&\lambda \Phi^+\Phi^- - (\Phi-\bar \Phi) (W^\RT +\lambda
\Phi\Phi^+\Phi^-) \, .
\eea
The matter fields $\Phi^\pm$ are oppositely charged under a U(1) symmetry.
The corresponding $D$-term enforces $|\Phi_+| = |\Phi_-|$ in the minimum.
By an overall phase rotation, $W_0$ can be made real and positive.
${\rm Im}(T)$ adjusts to minimise $W^\RT$.  Then there is an
overall phase between $W^\RT$ and the matter potential
$W^{\rm m}$.  The charged fields can be rotated to be real, and
the phase of $\lambda$ can be absorbed in $\Phi$. We can take
any residual phase dependence to reside in $\Phi$.  We thus define the
real fields
\be
T = X + \rmi Y \, , \qquad
\Phi = \phi + \rmi \alpha \, , \qquad \Phi^- = \Phi^+ = \phi^+ \, .
\label{var}
\ee

The U(1) symmetry is unbroken for a zero VEV of the charged fields
$\phi^+ =0$.  Cosmic strings form in a U(1) breaking phase
transition.  To see whether such a phase transition can take place
during or after inflation, we consider the stability of the potential for
$\phi^+=0$.   We will assume that moduli stabilisation is not
disrupted by the presence of the matter fields, and that the
moduli are fixed at some value $T_0$.  We will check this
assertion numerically.

The potential is indeed extremised for $\phi^+=0$, i.e.\
$\partial_{\phi^+} V|_{\phi^+ = 0} =0$. Whether this is a stable
minimum, a saddle or a maximum depends on the mass matrix.  The
mass matrix is block diagonal in the charged fields and the
neutral $\Phi$ field.  Let us start with the latter first. As a
consequence of the shift symmetry the $\phi$ direction is flat for
$\phi^+=0$.  The potential is extremised with respect to $\alpha$,
i.e.\ $\partial_\alpha V|_{\phi^+=0} = 0$, for $\alpha^2 = 0$ or
$\alpha^2 =\alpha_0^2 \equiv -1/2- V_\RT/(4\m^2)$.  The mass
of $\alpha$ at these extrema is
\be
m_\alpha^2 = \left \{
\begin{array}{ll}
2 \m^2 (2-y) & \quad \alpha =0 \\
-4 \m^2 \e^{2 \alpha_0^2} (2-y) &  \quad \alpha =\alpha_0
\end{array}
\right.
\label{malpha}
\ee
where we defined
\be y= - V_\RT /\m \, .
\label{y}
\ee
Here $V_\RT$ and $m = \e^{K^\RT/2} |W^\RT|$ are the
racetrack potential and gravitino mass as defined in the absence
of matter fields.  In particular $m \approx m_{3/2} \equiv
\e^{K/2} |W|$ up to small $\Phi_i$ dependent corrections, and with
abuse of language we will sometimes refer to it as the gravitino
mass. With $D$-term or D-brane uplifting, $y \approx 3$
after inflation and $\alpha = \alpha_0$ is the minimum.
However, during inflation $y \approx 1$ and $\alpha = 0$ is the
minimum. This implies that the phase field obtains a VEV during
inflation. As we will see, since the $\alpha$-field starts rolling
only near the very end of inflation,  it affects the inflationary
results only mildly (keep also in mind that \eref{malpha} is only
valid for unbroken U(1) with $\phi^+ =0$).

Consider now the bosonic mass matrix of the charged fields along the
$\phi$-flat direction.
The diagonal entries of the mass matrix are
\begin{equation}
m^2_{\Phi^+\bar \Phi^+}=m^2_{\Phi^-\bar \Phi^-}=
\e^{2\alpha^2} \Big( V_\RT +
(1+4\alpha^2)\m^2 + { \tilde \lambda^2 \vert \Phi\vert^2}
\Big) \, ,
\label{diag}
\end{equation}
and  the off-diagonal term reads
\begin{equation}
m^2_{\Phi^+\Phi^-}= \frac{\tilde{\lambda} \e^{2\alpha^2}}{(T+\bar
T)^{3/2}} \Big(2\rmi \alpha \bar W^\RT - (T+\bar T)
\partial_{\bar T} \bar W^\RT \Phi + \bar W^\RT (2 + 4\alpha^2)\Phi
\Big) \, . \label{offdiag}
\end{equation}
Here we defined the rescaled coupling via
\be
\tilde{\lambda} = \frac{\lambda}{(T+\bar T)^{3/2}} \, .
\ee
The bosonic mass eigenstates are $ m^2_{\pm} = m^2_{\Phi^+\bar
  \Phi^+}\pm \vert m^2_{\Phi^+\Phi^-}\vert$. The fermion mass
eigenstates are two Weyl fermions with masses 
$m^2_{\psi}= \e^{2\alpha^2}\tilde{\lambda}^2 \vert \Phi\vert^2$.

What does this imply for inflation?  Consider first the situation at
the beginning of inflation, near the racetrack saddle point.  As
discussed above $\alpha =0$, and the mass matrix elements simplify to 
\begin{equation}
m^2_{\Phi^+\bar\Phi^+}\approx (1-y)\m^2 + \tilde \lambda^2 \phi^2 \, , \qquad 
m^2_{\Phi^+\Phi^-}  \sim \tilde \lambda \phi \m \, .
\label{msaddle}
\end{equation}
The $\phi_\pm =0$ extremum is only stable for large $\phi$
\begin{equation}
\phi >  \phi_c \approx \frac{\m}{\tilde \lambda} (1+ \sqrt{4y-3})
\label{phic}
\end{equation}
assuring both bosonic mass eigenstates to be positive definite. In
terms of the original parameters in the potential ${\m}/{\tilde
\lambda} \sim {W_0}/{\lambda}$. In racetrack inflation $\m \sim
10^{-8}$ which gives $\phi_c \ll 1$.  If follows that the
resulting behaviour will be very different depending on whether
$\phi$ is smaller or larger than the critical value at the onset
of inflation.  The flat direction is lifted by loop corrections.
If these corrections give a large mass to the $\phi$ field,
$m_\phi > H_*$, the field will settle in the minimum, whereas in
the opposite limit a large VEV may be expected.

The loop corrections originate from the splitting between the
boson and fermion masses as a consequence of SUSY breaking. Notice
that ${\rm Str} M^2$ is $\phi$-independent (but does not vanish in
supergravity) so the $\phi$ potential arises from the log
correction
\bea
\fl V=\frac{1}{32\pi^2} \Bigg\{ & (m^2_{\Phi^+\bar \Phi^+}- \vert
m^2_{\Phi^+\Phi^-}\vert)^2 \ln \frac{(m^2_{\Phi^+\bar \Phi^+}-
\vert m^2_{\Phi^+\Phi^-}\vert)}{\Lambda^2}
\nonumber \\ \fl & {}
+ (m^2_{\Phi^+\bar
\Phi^+} + \vert m^2_{\Phi^+\Phi^-}\vert)^2 \ln
\frac{(m^2_{\Phi^+\bar \Phi^+  } +\vert
m^2_{\Phi^+\Phi^-}\vert)}{\Lambda^2} -2 m^4_\psi \ln
\frac{m_\psi^2}{\Lambda^2} \Bigg\} \, .
\eea
If $\phi$ (or $\alpha$) is very large the mass splitting, and thus
the potential, vanishes as in supersymmetry. Hence a far away D7 is
weakly attracted by the D3.  We can estimate the potential in the
limit that $\m^2 \ll \tilde \lambda^2 (\phi^2+\alpha^2)$:
\be
V \sim \frac{1}{32 \pi^2}
\Big\{ \tilde \lambda^2(\phi^2 + \alpha^2) \m^2+ \Or(\m^4) \Big\} \, .
\ee
Then $m_{\phi}^2 /H_*^2 \sim \tilde \lambda^2/(32\pi^2)$ is
indeed small unless $\tilde \lambda$ approaches non-perturbative
values. We therefore conclude that $\phi$ is light during inflation,
and remains frozen.  We generically expect $\phi> \phi_c $ at the
onset of inflation.  But since smaller  VEVs are not excluded, we
discuss both cases in turn.

\begin{figure}
\centerline{
\includegraphics[width=8cm]{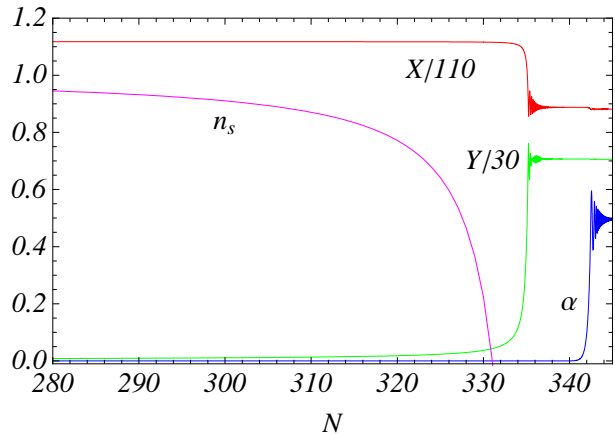}
} \caption{Evolution of the (rescaled) fields X, Y and $\alpha$ as
function of the number of $e$-folds $N$ since the beginning of
inflation, for the case $\phi > \phi_c$ and the U(1) symmetry is preserved.
The parameters are $A = 1/50,\, B = -7/200,\, a = \pi/50,\, b = \pi/45, \,
    W_0 = -3.5868 \times 10^{-5},\, \lambda = 1$.} \label{F:flds}
\end{figure}

\begin{figure}
\centerline{
\includegraphics[width=8cm]{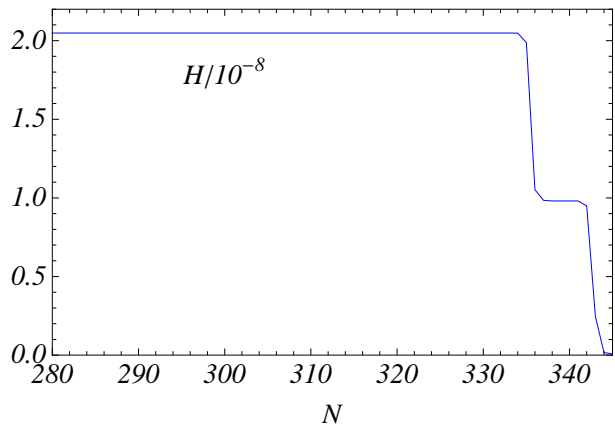}
} \caption{Evolution of the Hubble rate as a function of the
number of e-folds $N$. The parameters are the same as in
Figure~\ref{F:flds}.} 
\label{F:H}
\end{figure}

\subsection{Unbroken U(1)}
\begin{figure}
\centerline{
\includegraphics[width=8cm]{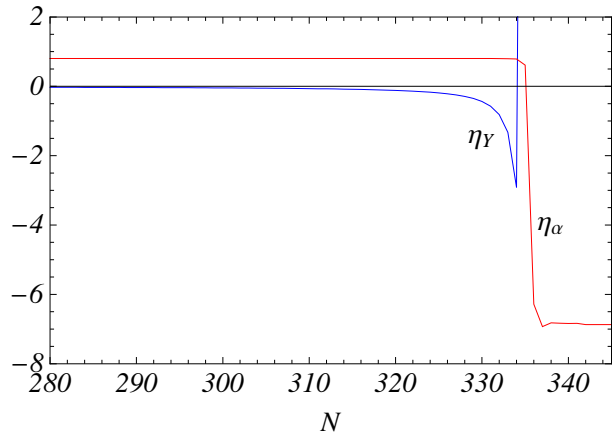}
} \caption{Plotted is $\eta_Y$ and $\eta_\alpha$ parameter during
inflation. Same parameters as in Figure~\ref{F:flds}.}
\label{F:eta}
\end{figure}

Consider first the case that $\phi > \phi_c$, and the U(1) preserving phase $\phi^+ =0$ is a minimum of the potential at the saddle. As discussed above the $\phi$ field is frozen during inflation. The evolution of the fields $X,Y,\alpha$ is shown in Figure~\ref{F:flds}. At the end of racetrack inflation, the phase field $\alpha$ becomes tachyonic and starts rolling towards the minimum
$\alpha=\alpha_0$ starting from $\alpha=0$ during racetrack
inflation, as follows from \eref{malpha}.   The potential near $\alpha = 0$ can be approximated as
\begin{equation}
V(\alpha)=V_0(1 +  \eta_\alpha \alpha^2) \, .
\end{equation}
where it should be noted that the canonically normalised field is
$\alpha/\sqrt{2}$, and that $\eta_\alpha <0$. This gives rise to a
period of fast-roll $\alpha$-driven inflation, which ends when
$\alpha$ departs from the origin too much and feels the minimum of the
potential.  Figure~\ref{F:H} shows the Hubble constant during
inflation, which is nearly constant both during racetrack and the
following period of fast-roll inflation.  Since $|\eta_\alpha| =
\Or(10)$ ($|\eta_\alpha| \approx 7$ for the parameters used in Figure
\ref{F:flds}) is rather large, this extra bout of inflation is
short. The number of extra e-folds can be approximated by
\cite{fastroll} 
\be
N \approx \frac1F \log\(\frac{\alpha_0}{\alpha_*} \) \, , \qquad
F = \sqrt{\frac{9}{4} + 3\eta_\alpha} - \frac{3}{2}
\ee
with $\alpha_0 \sim 0.5$ the value at the minimum, and $\alpha_*$ the
initial value.  During the period of racetrack inflation 
$\eta_\alpha \approx 1$ and the $\alpha$-field is rapidly damped to
zero. Thus we expect $\alpha_*$ to be small, set by the scale of quantum
fluctuations $\alpha_* \sim H_* \sim 10^{-8}$.  Plugging in the
numbers gives $N_{\rm fast \, roll} \sim 5$, in good agreement with the
numerical results shown in Figures~\ref{F:flds} -- \ref{F:eta}. This
extra burst of fast-roll inflation lowers the spectral index, as now
observable scales leave the horizon $N_* -5$ e-folds before the end of
the first period of racetrack inflation (instead of the usual 
$N_* \approx 55$).  But the change is minimal,  lowering $n_s$ by less than
$0.01$.

In the minimum after inflation $\alpha^2 = \alpha_0^2 \approx 1/4$.
The situation is now very 
different from the one at the saddle point.  Indeed, the diagonal
and off-diagonal terms of the bosonic mass matrix are now
\begin{equation}
m^2_{\Phi_+\bar\Phi_+}\approx ( -\m^2+ \tilde \lambda^2 (\alpha^2 + \phi^2)) 
\, , \qquad m^2_{\Phi+\Phi_-} \sim \tilde \lambda (\phi+
\alpha) \m \label{msad}
\end{equation}
where in the off-diagonal term we neglected order one
coefficients.  It is clear that for $\phi < \alpha$ the VEV of the
flat direction field $\phi$ will only affect the mass eigenstates
at subleading order.  The result is that the charged fields are
non-tachyonic for all values of $\phi$.  There is no equivalent of
$\phi_c$ as found at the saddle point.  The minimum is U(1)
preserving and stable.

To summarise, the U(1) matter symmetry is preserved to low
energies if at the onset of inflation $\phi>\phi_c$. Racetrack
inflation proceeds as before, but is followed by a short period of
$\alpha$-driven fast-roll inflation. This lowers the spectral
index a little bit. No cosmic strings are formed at any time.

\subsection{Broken U(1)}

\begin{figure} \centerline{
\includegraphics[width=8cm]{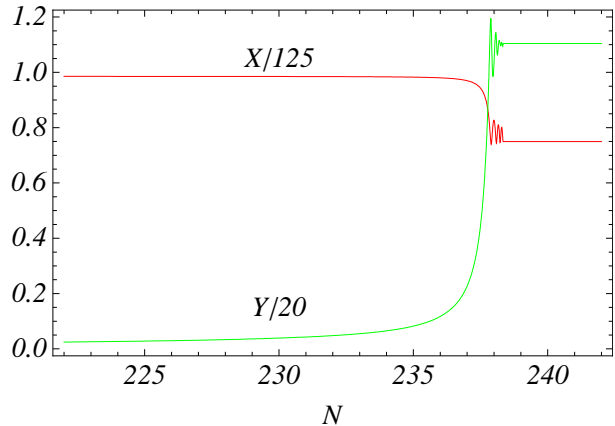}}
\caption{The evolution of the (rescaled) moduli fields $X$ and $Y$ during
inflation for the case $\phi<\phi_c$.  The parameters values are
$A = 1/50,\, B = -7/200,\, a = \pi/50,\, b = \pi/45,\,
W_0 = -1/25000 \times 10^{-5}, \,.\lambda = 1$.}
\label{F:flds2}
\end{figure}

\begin{figure}
\centerline{
\includegraphics[width=8cm]{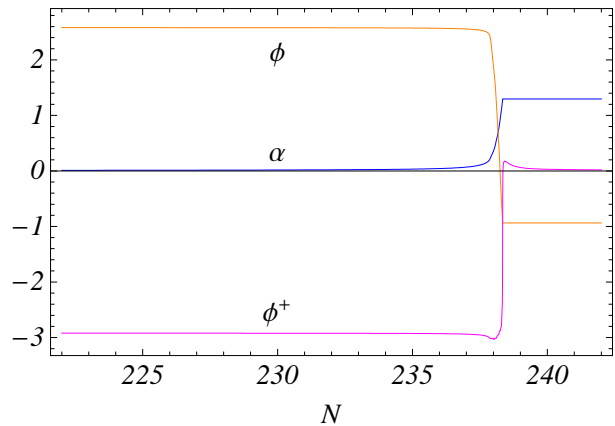}
} \caption{The evolution of the charged field $\phi^+$ and
$\alpha$ during inflation when $\phi<\phi_c$. The parameters values are the same
as in Figure~\ref{F:flds2} } \label{F:flds3}
\end{figure}

If the initial field value is small $\phi < \phi_c$  the charged
fields condense towards a $U(1)$ breaking minimum with $\Phi$ and
$\phi^+ = \phi_-$ non-zero.   From  the mass matrix
(\ref{diag},~\ref{offdiag}) it follows that the mass eigenstates
approach $(1-y)\m$ in this limit.  Since $y$ is slightly bigger than
one at the saddle (extracted from the numerics), this gives negative
mass eigenstates, and $\eta_{\phi^+} \sim (1-y) = -\Or(0.1)$.
Inflation still proceeds, but the U(1) is broken both during and after
inflation.  Any strings formed in the U(1) breaking phase transition
are inflated away. Figures \ref{F:flds2} and \ref{F:flds3} show the
evolution of the various fields during inflation.

The minimum after inflation can be found in the small field limit, by
solving $F_{\Phi_i} = 0$.  The result is that all matter fields
$\phi,\alpha,\phi^\pm$ obtain a VEV of order $\m/\tilde\lambda$.  The
system thus ends up in a different minimum depending on whether $\phi$
larger or smaller than $\phi_c$ initially.  The U(1) preserving
minimum with $\phi^\pm =0$ is lower than the U(1) broken minimum
discussed in this subsection, by an amount $\delta V \sim \m^2$.  The
two minima are separated by a barrier.

\subsection{$F$-term lifting}

Up to now we have discussed a set-up where lifting was done by either
$D$-terms or anti-D-branes, and $y > 1$ during and after inflation
(see \eref{y}).  Depending on the initial conditions, we found that
either the U(1) symmetry is never broken, or else breaking occurs
before the onset of inflation and strings are inflated away.  Since
the $y$ value determines the $\alpha$-mass, and thus whether there is
a critical value $\phi_c$ or not, it may be interesting to consider
$F$-term uplifting as well as it gives $y = 0$ in the vacuum (see
e.g.~\cite{fterm} and references therein for $F$-term
uplifting). Although no explicit model of racetrack inflation with
$F$-term lifting exists, considering its robustness we expect it to be
possible.

With $F$-term lifting $\alpha =0$ is the minimum during and after
inflation, and the mass matrix for the charge fields is well
approximated by \eref{msaddle} throughout.  One may be inclined to
think that as $\phi$ drops below its critical value a U(1) breaking
phase transition takes place.  But this is not the case.  As $y$ drops
below $3/4$, which is the case in the post-inflationary minimum where
$y =0$, the charged fields become non-tachyonic for all
$\phi$-values. This is reflected in \eref{phic} by the fact that
$\phi_c$ becomes imaginary, i.e.\ non-existent.

The situation is thus similar to that for the $D$-term or D-brane
lifting case.  If $\phi > \phi_c$ initially the U(1) symmetry is
preserved throughout, and no strings form.  But since $\alpha =0$ both
during and after inflation, there is no period of $\alpha$-driven
fast-roll inflation.  In the opposite limit $\phi < \phi_c$, the
matter U(1) is broken throughout, and strings are inflated away.

\section{Discussion}
\label{s:discussion}

In this paper we discussed racetrack inflation coupled gravitationally
to a matter sector with a U(1) symmetry.  We found that cosmic
strings, if formed at all, do not survive to low energies. But how
generic are these results?  After all, we assumed a matter sector with
a shift symmetry for the neutral field, and we set  possible FI terms
to zero.  Let us comment on possible modifications to this scenario.

First of all, we could have introduced a FI term in the $D$-term for
the matter fields, i.e.\ we could have made the U(1)
pseudo-anomalous. This can be implemented by making the modulus field
transform under the U(1) (the U(1) transformation acts as a shift
symmetry on $T$) \cite{FI}.  To preserve gauge invariance of the
moduli superpotential, additional charged meson fields are needed
\cite{Dterm2}.  The upshot of all of this is that if the modulus and
the meson fields are stabilised at some non-zero value, this generates
an effective FI term for the matter fields.  If the same $D$-term is
used for uplifting, the model does not work as the charged fields will
cancel the $D$-term and therefore remove the uplifting.  On the other hand
additional lifting terms could be introduced. For the sake of
argument, let us introduce an anti-brane for uplifting, which does not
depend on the matter fields.   The $D$-term enforces $\Phi^-$ zero
throughout. With a shift symmetry for the $\Phi$ field the $F$-term
potential in the U(1) phase with $\phi^+ = 0$ is the same as the one
we have investigated. The  $\alpha$ minimum is also the same . The
mass eigenvalues for the charged fields are now 
$m_{\Phi^+ \Phi^+} \pm \sqrt{|m_{\Phi^+ \Phi^-}|^2 + g^2 \xi}$. 
The gauge coupling $g$ and
FI term $\xi$ depend on the moduli fields.  The $F$-term contribution
to the mass terms is the same as for $\xi =0$, and still given by
(\ref{diag},\ref{offdiag}). As a result,  the whole discussion of the
previous section follows through, although with slightly altered
parameters. Therefore depending on initial values, the U(1) stays either
broken or preserved, both during and after inflation.

The shift symmetry is an essential part of the set-up we have
discussed. Without it there is no flat direction.  Performing  the
same analysis again  with a canonical K\"ahler potential leads to
the result that the U(1) symmetry is preserved to low energies
for $\lambda \gtrsim W_0 \sim 10^{-5}$.  The matter fields are
minimised at $\phi^\pm =0$ and $\phi \approx \sqrt{2}$.  Racetrack
inflation proceeds as before, with the same predictions for the power
spectrum (of course the parameter values needed to tune $\eta \ll 1$
and to get a Minkowski minimum are slightly altered due to the presence of the
matter fields).

It is less clear what happens in the small coupling limit $\lambda <
W_0$, or equivalently $\tilde \lambda < m$.  The situation is similar
for all set-ups, with or without a shift symmetry or FI term.  We find
that the U(1) symmetry is broken.  For the set-up discussed in section
4, this  can be seen from \eref{phic}, as for such small couplings the
critical value $\phi_c$ is pushed above unity. The VEVs for the matter
fields are large.  In the small field limit we found 
$\phi^\pm, \phi,\alpha \sim W_0 /\lambda$. Extrapolating, we expect
field values of order one. As a result racetrack inflation is greatly affected.
Numerically, we did not find a working model with $Y$ the only
unstable direction at the saddle point.  This does not mean there is
not such a saddle as the potential is complex with many extrema.  It
does show that if racetrack inflation works the parameters and field values are
rather different from the case without the matter fields.  This is
rather surprising as in the $\lambda \to 0$ limit one expects the
effects of the matter fields to be small.  Although in this limit $W$
no longer depend on $\Phi_i$, but the K\"ahler potential and thus
the scalar potential does. All in all, this case leads to large
modifications of racetrack inflation, possibly even destroying it.

To conclude, in this paper we studied racetrack inflation
augmented by a matter sector.  The coupling between the moduli
fields and the matter fields is only gravitational.  The matter
sector is inspired by a D3/D7 system, and consists of a neutral
field $\Phi$ and two fields $\Phi^\pm$ with opposite charges under
a U(1) symmetry; the moduli fields are neutral under this U(1).
Due to a shift symmetry the $\phi ={\rm Re} (\Phi)$ direction is
flat.  The matter superpotential is $W^{\rm m} = \lambda \Phi
\Phi^+ \Phi^-$.  We investigated whether the matter sector can
affect inflation, and vice versa, and whether the moduli sector can
induce interesting effects in the matter sector such as cosmic
string formation.

The resulting scalar potential has both a U(1) preserving and breaking minimum,
the symmetry breaking induced by supergravity effects due to the presence of
the moduli sector.  One may be led to think this can give rise to cosmic
string formation.  We investigated this possibility in this paper. The
inflationary dynamics depend on initial conditions.  In particular if
$\phi$ is larger than some critical value at the onset of inflation, then the
U(1) symmetry is preserved both during and after inflation. If lifting is done
by $D$-terms or D-branes (and not by $F$-terms) racetrack
inflation proceeds as before, but is followed by a short period of
$\alpha$-driven fast-roll inflation. This lowers the spectral index by
a small amount, less than $0.01$. No cosmic strings are formed at
any time.  In the opposite limit that $\phi$ is small initially, the
U(1) symmetry is broken at all times.  Any strings formed at early
times are inflated away.

\ack
MP acknowledges the support of the "Impuls- und Vernetzungsfond"
of the Helmholtz Association, contract number VH-NG-006. CvdB and ACD are
supported in part by STFC.


\end{document}